\def\kms{\ifmmode{\,\hbox{km}\,s^{-1}}\else {\rm\,km\,s$^{-1}$}\fi}

\def\kmsm{{\rm\,km\,s^{-1}\,Mpc^{-1}}}

\def\hmpc{\ifmmode{h^{-1}\,\hbox{Mpc}}\else{$h^{-1}$\thinspace Mpc}\fi}
\def\hkpc{\ifmmode{\,h^{-1}\,{\rm kpc}}\else {$h^{-1}$\,kpc}\fi}

\documentstyle[11pt,paspconf]{article}

\begin{document}

\title{Galaxy Clustering 
	in the CNOC2 Redshift Survey}

\author{R.~G.~Carlberg\altaffilmark{1,2},
H.~K.~C.~Yee\altaffilmark{1,2},
S.~L.~Morris\altaffilmark{1,3},
H.~Lin\altaffilmark{1,2,4,5}, \\
P.~Hall\altaffilmark{1,2},
D.~Patton\altaffilmark{1,6},
M.~Sawicki\altaffilmark{1,2,7}, 
and
C.~W.~Shepherd\altaffilmark{1,2}
}

\altaffiltext{1}{Visiting Astronomer, Canada--France--Hawaii Telescope, 
        which is operated by the National Research Council of Canada,
        le Centre National de Recherche Scientifique, and the University 
	of Hawaii.}
\altaffiltext{2}{Department of Astronomy, University of Toronto, 
        Toronto ON, M5S~3H8 Canada}
\altaffiltext{3}{Dominion Astrophysical Observatory, 
        Herzberg Institute of Astrophysics,    ,  
        National Research Council of Canada,
        5071 West Saanich Road,
        Victoria, BC, V8X~4M6, Canada}
\altaffiltext{4}{Steward Observatory, University of Arizona,
        Tucson, AZ, 85721}
\altaffiltext{5}{Hubble Fellow}
\altaffiltext{6}{Department of Physics \& Astronomy,
        University of Victoria,
        Victoria, BC, V8W~3P6, Canada}
\altaffiltext{7}{Mail Code 320-47, Caltech, Pasadena 91125, USA}

\begin{abstract} 
The correlation evolution of a high luminosity subsample of the CNOC2
redshift survey is examined.  The sample is restricted to galaxies for
which the k corrected and evolution corrected R luminosity is $M_R \le
-20$ mag, where $M_\ast\simeq -20.3$ mag. This subsample contains
about 2300 galaxies.  In consort with 13000 galaxies in a similarly
defined low redshift sample from the Las Campanas Redshift survey we
find that the comoving correlation can be described as $\xi(r|z) =
(r_{00}/r)^\gamma (1+z)^{-(3+\epsilon)}$ with
$r_{00}=5.08\pm0.08\hmpc$, $\epsilon=0.02\pm 0.23$ and
$\gamma=1.81\pm0.03$ over the $z=0.03$ to 0.65 redshift range in a
cosmology with $\Omega_M=0.2$, $\Lambda=0$. The measured clustering
amplitude, and its evolution, are dependent on the adopted cosmology.
The evolution rates for $\Omega_M=1$ and flat low density models are
$\epsilon=0.9\pm0.3$ and $\epsilon=-0.5\pm0.2$, respectively, with
$r_{00}\simeq 5\hmpc$ in all cases.
\end{abstract}


\section{Introduction}

The measurement of the evolution of galaxy clustering is a direct test
of theories for the formation of structure and galaxies in the
universe.  Within CDM-style structure formation theories galaxies
form exclusively within dark matter halos whose clustering
evolution should be slower than the underlying dark matter.
Clustering evolution is of empirical and pragmatic interest
on scales comparable to the size of galaxies themselves, since
clustering leads to galaxy-galaxy merging of gas and stars which is a
physical driver of galaxy evolution. On somewhat larger scales galaxy
groups and clusters contain hot gas which exerts evolution pressures
on galaxies.

The theoretical groundwork to interpret the quantitative evolution of
dark matter clustering and the qualitative trends of galaxy clustering
evolution is largely in place for hierarchical structure models.
Although the details of the mass buildup of galaxies and the evolution
of their emitted light are far from certain at this time, clustering
of galaxies depends primarily on the distribution of initial density
fluctuations on the mass scale of galaxies.  N-body simulations of
ever growing precision and their theoretical analysis
(\cite{DEFW,EDWF}) have lead to a good semi-analytic understanding of
dark matter clustering into the nonlinear regime.  Normal galaxies,
which are known to exist near the centers of dark matter halos with
velocity dispersions in the approximate range of 50 to 250 $\kms$,
will not have a clustering evolution identical to the full dark matter
density field (\cite{kaiser84}).

The observational measurement of clustering at higher redshifts is
still maturing, with no published survey of the size, scale coverage,
or redshift precision of the pioneering low redshift CfA survey
(\cite{dp}).  There are two redshift surveys extending out to $z\simeq
1$, the Canada France Redshift Survey (\cite{cfrs_lf}) and the Hawaii
K survey (\cite{cowie}). Their redshift precision is insufficient for
kinematic studies, however both quantify the substantial evolution in
the luminosity function considerable population evolution. Measurement
of the correlation evolution of the galaxies in these surveys found a
fairly rapid decline in clustering with redshift (\cite{cfrsxi,kkeck})
although neither analysis took into account the evolution of the
luminosity function or was able to quantify the effects of the small
survey volumes.

\section{The CNOC2 Sample}

The Canadian Network for Observational Cosmology field galaxy redshift
survey (CNOC2) is designed to investigate nonlinear clustering
dynamics and its relation to galaxy evolution on scales smaller than
approximately 20\hmpc\ over the $0.1\le z \le 0.7$ range.  There is
substantial galaxy evolution over this redshift range
(\cite{bes,ldss,cfrs_lf,cowie,huan_lf}).  Investigating the clustering
evolution of a population requires some basic ability to recover its
progenitor hosts at higher redshift.  

\begin{figure}
\plotfiddle{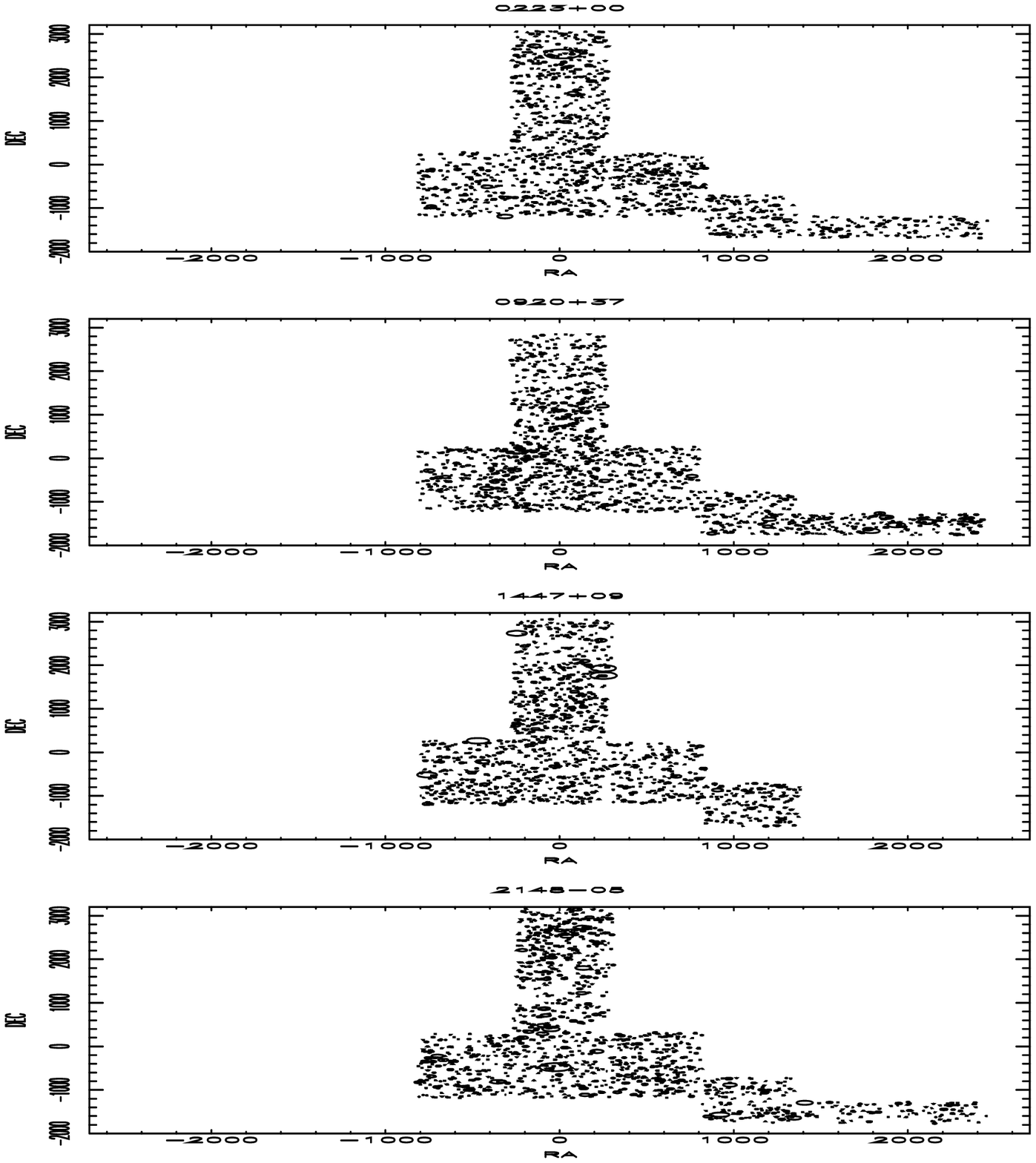}{12truecm}{0.0}{20}{80}{-190}{-300}
\plotfiddle{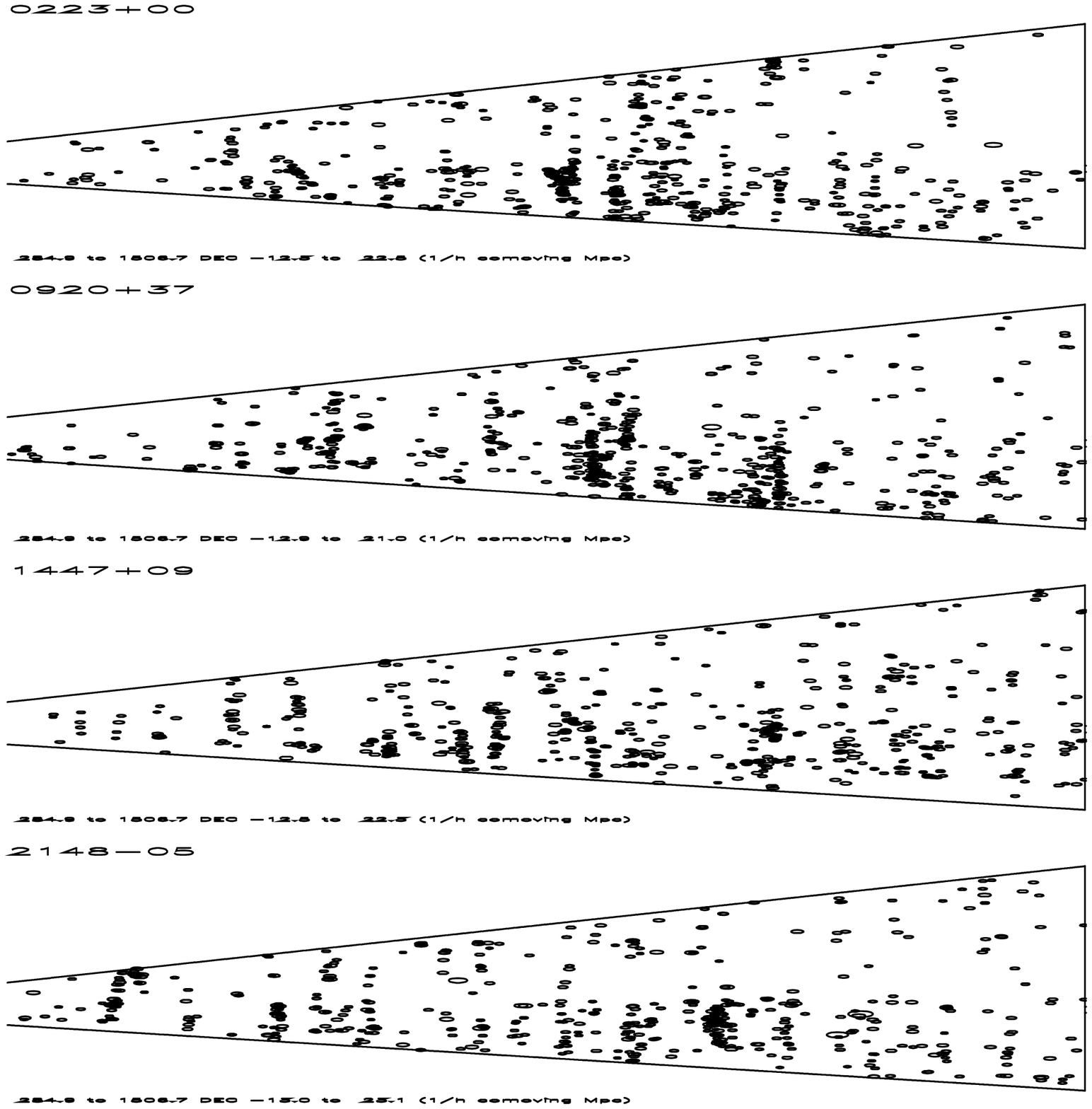}{12truecm}{0.0}{40}{80}{-60}{60}
\vspace{-7.0truecm}
\caption{The layout of the high luminosity redshift sample,
$M_R^{ke}\le -20$ mag, sample on the sky (left) and in co-moving
distance (redshifts from 0.1 to 0.65) versus declination in comoving
coordinates (right).
\label{fig:xyz}}
\end{figure}

The observing and catalogue design procedures are adaptations and
extensions of those for the CNOC cluster survey (\cite{yec,yee_iau}).
The CNOC2 survey is contained in four patches on the sky subtending a
total sky area of about 1.55 square degrees.  The sampled volume is
about $0.5\times 10^6 h^{-3} {\rm Mpc}^3$, roughly comparable to the
low redshift ``CfA'' survey used for similar measurements at low
redshift (\cite{dp}) which had 1230 galaxies in the ``semi-volume
limited'' Northern sample from which the correlation length was
derived.  The CNOC2 limiting magnitude of $R=21.5$ magnitude
efficiently covers the redshift range targeted, 0.1 to about 0.7.  The
survey layout for the high luminosity subsample analyzed here is shown
in Figure~\ref{fig:xyz}.

Photometry is obtained in the UBVRI bands, with the R band fixing the
sample limit at 21.5 mag. The R filter stays redward of the 4000\AA\
break over our redshift range.  The other bands provide
information useful for determining appropriate k-corrections and
separating galaxies into types of different evolutionary state (
an issue not considered in this paper). 

At low redshift we use the Las Campanas Redshift Survey (LCRS) to
provide a directly comparable sample. The LCRS is an R band selected
survey (\cite{lcrs}) that covers the redshift range 0.033 to 0.15,
with R band flux between 15.0 and 17.7 mag.  The same correlation
analysis programs were used with for the LCRS and CNOC2 data. The only
differences are that the LCRS data are not k corrected and is only
analyzed for a single cosmological model, $q_0=0.1$.

\subsection{The High Luminosity Subsample}

The evolution of correlations with redshift is sensitive to luminosity
evolution.  That is, usually the goal is to compare the evolution of
the correlations of the same ``primary'' galaxies, which have likely
evolved in luminosity as a result of stellar evolution of the existing
stars, star formation, and merging of companion galaxies.  Over the
redshift range discussed here these effects are significant, but
basically perturbative on a stable underlying population. In the
absence of a luminosity correction one will be adding in galaxies of
lower intrinsic mass with increasing redshift. Lower luminosity
galaxies are less correlated (\cite{roysoc,loveday}) which if
over-represented in the higher redshift will lead to an erroneously
rapid correlation evolution. From previous studies we concluded that
the luminosity function can be approximated as evolving such that the
absolute magnitude varies as $M(z) = M(0)-Q z$ with $Q\simeq 1$
(\cite{huan_lf}).  It should be noted that the this approach is a
statistical correction designed to identify the same population at two
redshifts and will work even for an evolving rate of bursting star
formation, provided that evolution has no significant mass
dependence. This assumption should be adequate for our high luminosity
subsample (\cite{cfrs_lf,huan_lf}) but may fail for at lower
luminosities.  The volume density of galaxies makes no difference to
the correlations.  For the correlation analysis here we will use
galaxies with $M_R^{k,e}
\le -20$ mag, which defines a volume limited sample over the $0.1\le z
\le 0.65$ range. The resulting sample (for $q_=0.1$) contains 2285
galaxies.

The LCRS data are evolution corrected with the same $Q$ as the CNOC2
data, although at a mean redshift of about 0.1, this makes very little
difference. The resulting low redshift sample contains 12467 galaxies
that are used in the correlation analysis.

\section{Real Space Correlations}

The CNOC2 sample is designed to measure nonlinear clustering on scales
of 10\hmpc\ and less. The clustering is quite naturally measured in
terms of the two-point correlation function, $\xi(r)$, which measures
the galaxy density excess above a random distribution, $n_0$. at
distance $r$ from a galaxy, $n(r) = n_0[1+\xi(r)]$ (\cite{lss}).
Measurement of the real space function $\xi(r)$ is not straightforward
with redshift space data in any case, and yet more difficult with only
a few hundred galaxies per redshift bin. The projected real space
correlation function removes the peculiar velocities of redshift space
at the cost of making a choice for the length of the redshift column
over which the summation is done.  Noting that the correlation
function measures the fractional variance in galaxies in different
volumes we check as much as possible that our evaluation of the
correlation does not artificially add or remove variance which will bias
the result.

\subsection{The Projected Real Space Correlation Function}

The correlation function is a real space quantity, whereas the
redshift space separation of two galaxies depends on their peculiar
velocities as well as the physical separation.  Although the peculiar
velocities contain much useful information about clustering dynamics
it is mainly a complication for the study of configuration space
correlations. The peculiar velocities are eliminated by integrating
over the redshift direction to give the projected correlation
function,
\begin{equation}
w_p(r_p) = \int_{-R_p}^{R_p} \xi(\sqrt{r_p^2+r_z^2})\,dr_z
\label{eq:wp}
\end{equation}
(\cite{dp}). If $R_p=\infty$ and $\xi(r)$ is a power law,
$\xi(r)=(r_0/r)^\gamma$, then Equation~\ref{eq:wp} integrates to $w_p(r_p)/r_p=
\Gamma(1/2)\Gamma((\gamma-1)/2)/\Gamma(\gamma/2)(r_0/r_p)^{\gamma}$
(\cite{lss}). However, in a practical survey summing over ever the
entire redshift extent of the survey leads to little increase in the
signal and growing noise from fluctuations in the field density.  Here
we are focussed on the non-linear correlations, $\xi>1$ which suggests
we use $R_p$ of several correlation lengths. A range of $R_p$ and the
resulting errors in $r_{00}$ (see Equation~\ref{eq:emodel} below) are
displayed in Figure~\ref{fig:rR}. We select $R_p=10\hmpc$ as a
conservative choice, having the largest (and most stable)
errors. Similar results are obtained for any $10\hmpc \le R_p\le
50\hmpc$, but with increasing statistical fluctuations, most notably
in the estimated errors of the fitted correlations.

\begin{figure}[ht]
\plotfiddle{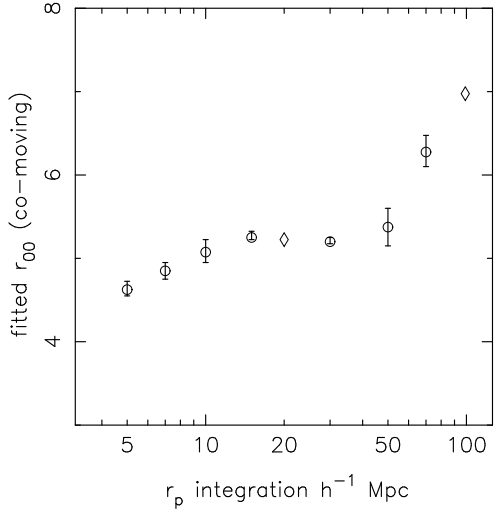}{8truecm}{0.0}{40}{40}{-120}{0}
\vspace{-0.5truecm}
\caption{The derived $r_{00}$ (see Eq.~\ref{eq:emodel}) 
as a function of the integration
length, $R_p$ used to define $w_p(r_p)$. The 90\% confidence intervals
are shown.  The diamonds are plotted when $\chi^2$ exceeds the formal
level for 90\% confidence. This most likely arises because the
variances used to calculate $\chi^2$ are estimated from the dispersion
of the four patches, which will sometimes lead to erroneously small
variances and hence large $\chi^2$ values.
\label{fig:rR}}
\end{figure}  

The choice of a statistical estimator of the correlation function
depends on the application.  With point data the basic procedure is to
determine the average number of neighboring galaxies within some
projected radius and redshift distance $R_p$.  We estimate $w_p(r_p)$
using the simplest estimator, $w_p =  DD/ DR -1$ (\cite{lss}), which is
accurate for the nonlinear clustering examined here and faster than
methods which include the RR sum.  The DD and DR sums extend over all
four patches in CNOC2 and the six slices of LCRS, so that patch to
path variations in the mean volume density become part of the
correlation signal. The patch to patch variation is used to estimate
the error.

Estimated projected correlation functions, in co-moving co-ordinates
using $R_p=10\hmpc$, are calculated for the LCRS galaxies bounded by
redshifts [0.033, 0.15] and six redshift bins for the CNOC2 data,
[0.10, 0.20, 0.26, 0.35, 0.40, 0.45, 0.55, 0.65].  The first CNOC2
redshift bin is the least populated with 185 galaxies and the fourth
bin has the most with 602 galaxies. Relatively narrow redshift bins
helps to reduce any problems associated with the detailed shape of the
variation of $n(z)$ over the redshift range of the bin.  To fit the
data to a specified function requires error estimates at each point.

\begin{figure}
\plotfiddle{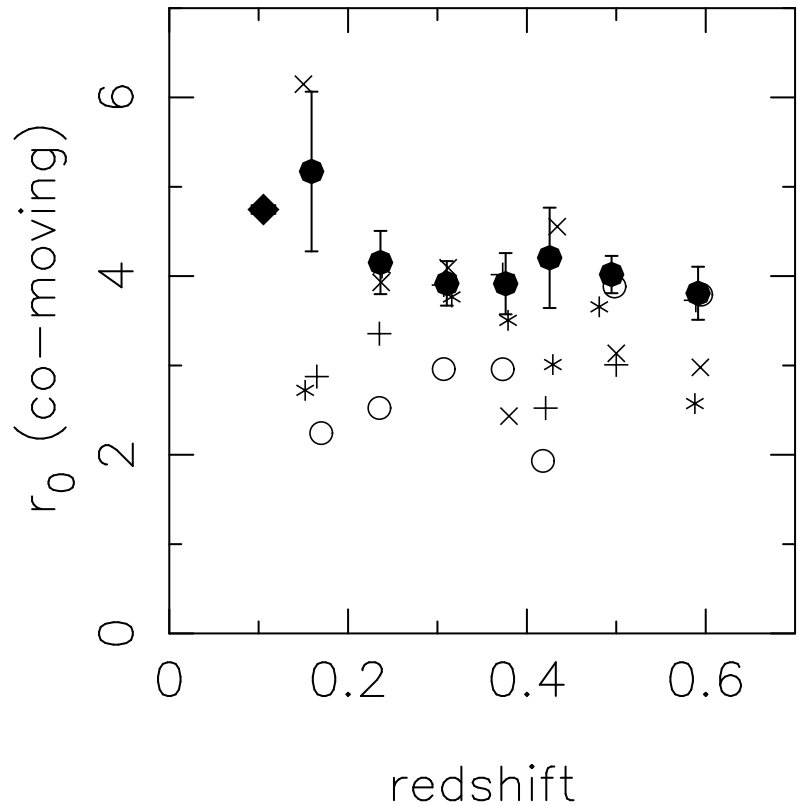}{12truecm}{0.0}{70}{70}{-200}{0}
\vspace{-1.0truecm}
\caption{The correlation lengths (normalized to $\gamma=1.8$) as a 
function of redshift for $\Omega_M=0.2,\Omega_\Lambda=0$. The darkened
diamond is from the LCRS with $q_=0.1$. The errors are estimated from
the variance of the four CNOC2 sky patches (shown with plus, asterisk,
circle and cross symbols for the 0223, 0920, 1417 and 2148 patches,
respectively) and the six LCRS slices (not shown since they are
small). The solid points are the result of the combined correlation
analysis of the four fields and are in general always larger than the
mean of the individual fields, since they include the field to field
variance.
\label{fig:r0z}}
\end{figure}

The correlations are fit to the projection of the power law
correlation function, $\xi(r)=(r_0/r)^\gamma$ taking the errors to be
the $1/\sqrt{DD}$, the weighted counts in each $r_p$ bin.  The fits
use only the $0.16\le r_p \le 5.0\hmpc$ range where there are minimal
complications from slit crowding and small correlations.  All results
here are derived using co-moving co-ordinates, and normalized to a
Hubble constant $H_0=$ $100 h \kmsm$. The results displayed in
Figure~\ref{fig:r0z} are derived assuming a background cosmology of
$\Omega_M=0.2,
\Omega_\Lambda=0$.

\subsection{Errors of the Correlations}

A straightforward approach to random error estimates is to take
advantage of our sample being distributed over a number of separate
patches.  In each patch we fit four for CNOC2 and six for LCRS.  The
open circles are the results for the four individual patches CNOC2
fields (the individual LCRS slices are so similar that they are not
displayed).  The solid points give the result from the combined data,
along with the estimated error.

As the size of a correlation survey grows there is a systematic
increase in the derived correlation length. This is is illustrated in
Figure~\ref{fig:r0z}.  The straight mean of the CNOC2 $r_0$ is
3.2\hmpc, the median is 3.4\hmpc, and the mean of $r_0^\gamma$ is
3.5\hmpc, whereas the four patches analyzed together find $r_0$ is
nearly 4.0\hmpc.  This raises the question as to whether the
correlations have converged within the current survey. We note that
the expected variation from patch to patch for the given volumes with
narrow redshift bins is about 45\%, which is consistent the difference
between a correlation length of 3.5 and 4.3\hmpc. In the combined
sample with bigger bins we expect that there could be as much about
10\% of the variance missing, which would boost the correlation
lengths another 5\%.

\section{The Evolution of Galaxy Clustering}

The evolution of correlations is conveniently described with the
``epsilon'' model,
\begin{equation}
r_0(z) = r_{00} (1+z)^{-(3+\epsilon-\gamma)/\gamma},
\label{eq:emodel}
\end{equation}
where all lengths are in co-moving units. The parameter $\epsilon$
measures the rate of growth of the mean physical density of
neighboring galaxies. If $\epsilon=0$ then there is no change in the
{\em physical} density with redshift.  Positive $\epsilon$ indicate a
decline of clustered density with increasing redshift. The derived
$r_0(z)$ along with the estimated errors, $\sigma(r_0(z))$ are fit to
the $\epsilon$ model with a formal $\chi^2$ which reports both
the suitability of the model and the parameter confidence interval.

\begin{figure}
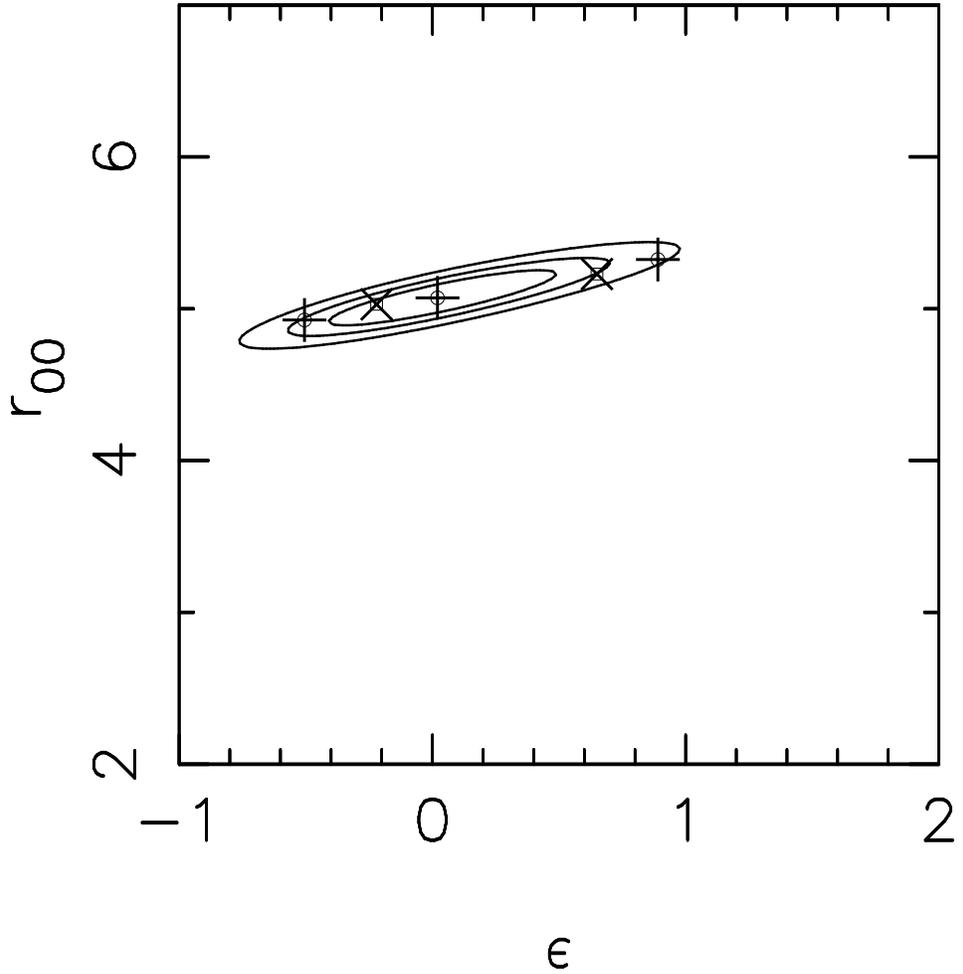

\plotfiddle{carlbergr_4.eps}{15truecm}{0.0}{85}{85}{-250}{14.5}
\plotfiddle{carlbergr_5.eps}{15truecm}{0.0}{85}{85}{-250}{453}
\plotfiddle{carlbergr_6.eps}{15truecm}{0.0}{85}{85}{-250}{891.5}
\plotfiddle{carlbergr_7.eps}{15truecm}{0.0}{85}{85}{-250}{1330}
\plotfiddle{carlbergr_8.eps}{15truecm}{0.0}{85}{85}{-250}{1768.5}

\vspace{-63truecm}
\caption{
The $\chi^2$ confidence levels for fits to the $\epsilon$ model for
the $\Omega_M=0.2,\Omega_\Lambda=0$ model. The contours are for 68\%,
90 and 99\% confidence. The plus signs
mark the results for $\Omega_M=1, \Omega_\Lambda=0$, peaking at
$\epsilon\simeq 0.6$ and $\Omega_M=0.2, \Omega_\Lambda=0.8$ peaking at
$\epsilon=-0.4$. The x's show the results for no evolution
correction, $Q=0$, in which case $\epsilon=0.65$, and
$Q=2$, evolution, where $\epsilon=-0.3$.
\label{fig:r00e}}
\end{figure}  

The correlation lengths for the CNOC2 and LCRS analyzed in precisely
the same way for our standard $R_p=10\hmpc$ and $Q=1$ are shown in
Figure~\ref{fig:r0z}. It is immediately clear that there is
relatively little correlation evolution of this sample. It must be
borne in mind that the sample is defined to be a similar set of
galaxies with luminosities larger than about $L_\ast$, with luminosity
evolution corrected, that accurately approximates a sample of fixed
stellar mass with redshift. Samples which admit lower luminosity
galaxies, or do not correct for evolution, or are selected in bluer
pass bands where evolutionary effects are larger and less certainly
corrected, will all tend to have lower correlation amplitudes.

The $\epsilon$ model fits to the measured correlations are shown in
Figure~\ref{fig:r00e}. The results are clearly a strong function of
the cosmological world model. However, the strength of the
correlations is truly impressive for any of the cosmologies. In the
case of the low density flat model, the distances to the highest
redshift galaxies are sufficiently large that the mean luminosity of
the highest redshift galaxies is larger than those at lower redshifts.
The best fit $\epsilon$ values are $0.02\pm 0.23$ for
$\Omega_M=0.2,\Omega_\Lambda=0$,

At redshifts beyond 0.1 or so, the choice of cosmological model has a
substantial effect on the correlation estimates. Relative to high
matter density cosmological models, Low density and $\Lambda$ models
tend to have larger distances and volumes, which causes the
correlations to be enhanced. The LCRS data are analyzed only within
the $q_0=0.1$ model. The correlations for three cosmologies, flat
matter dominated, open, and low density $\Lambda$, are shown in
Figure~\ref{fig:r0z}. The evolution rates for the flat and flat low
density models are $\epsilon=0.9\pm0.3$ and $\epsilon=-0.5\pm0.2$,
respectively, with $r_{00}\simeq 5\hmpc$ in all cases. These 
are marked with plus signs in Figure~\ref{fig:r00e}.

The effects of alternate values for the luminosity evolution are shown
in Figure~\ref{fig:r00e} with crosses indicating the results for $Q=0$
and $Q=2$, with the adopted value being $Q=1$. The absolute magnitude
limit remains $M_R=-20$ mag in all cases.  In the absence of any
allowance for luminosity evolution, $Q=0$, galaxies of lower current
epoch luminosity are increasingly included with redshift. Since lower
luminosity galaxies are generally less correlated that leads to an
artificially large $\epsilon$, the effect over this redshift range
being very roughly $\Delta \epsilon \approx 0.5 \Delta Q$.

\section{Conclusions}

The primary conclusion is that correlations evolve very weakly, it at
all, with redshift. The main correlation results of this paper are
contained in Figures~\ref{fig:r0z}. Out to redshift 0.65 the
correlation evolution of high luminosity galaxies, k and evolution
corrected, can be described with the double power law model, $\xi(r|z)
= (r_{00}/r)^\gamma (1+z)^{-(3+\epsilon)}$, with
$r_{00}=5.08\pm0.08\hmpc$, $\epsilon=0.02\pm 0.23$ and
$\gamma=1.81\pm0.03$. There is no evidence in the current data for a
change in the slope of the correlation function with redshift.  The
clustering results are consistent with galaxies being located in dark
matter halos (\cite{ccc,baugh,ckkk}) but are also consistent with
``light-traces-mass'' in low density universes where the mass
clustering evolves very slowly with redshift.

Environmental factors clearly play a major role in the development of
galaxies, as most clearly seen in the morphology-density relation
(\cite{dressler,hashimoto}). We have established here that the
neighborhood density, at separations between 0.l and about 5\hmpc, is
essentially non-evolving with increasing redshift. The implication is
that any evolution in the population cannot be attributed to a
significant change in the environmental density of neighboring
galaxies.  In a flat low-density model there is a slight rise of the
mean density of neighbors into the past. The high luminosity luminous
galaxies examined here do not as a population evolve much over
the $z\le 1$ range (\cite{cfrs_lf,cowie,huan_lf}) which could be the
outcome of either no change in the environment, or, these galaxies
being relatively insensitive to environmental change.

\acknowledgments

This research was supported by NSERC and NRC of Canada. We thank the
CFHT Corporation for support, and particularly the telescope operators
for their enthusiastic and efficient control of the telescope.
HL acknowledges support provided by NASA through Hubble Fellowship grant
\#HF-01110.01-98A awarded by the Space Telescope Science Institute, which 
is operated by the Association of Universities for Research in Astronomy, 
Inc., for NASA under contract NAS 5-26555.

\end{document}